\documentclass[sn-chicago,Numbered]{sn-jnl}
\usepackage{graphicx}%
\usepackage{multirow}%
\usepackage{amsmath,amssymb,amsfonts}%
\usepackage{amsthm}%
\usepackage{mathrsfs}%
\usepackage[title]{appendix}%
\usepackage{xcolor}%
\usepackage{textcomp}%
\usepackage{manyfoot}%
\usepackage{booktabs}%
\usepackage{algorithm}%
\usepackage{algorithmicx}%
\usepackage{algpseudocode}%
\usepackage{listings}%
\theoremstyle{thmstyleone}%
\theoremstyle{thmstyletwo}%
\theoremstyle{thmstylethree}%
\newtheorem{definition}{Definition}%

\begin{document}

\title[Article Title]{Comment on ``Improved RSA Technique with Efficient Data Integrity Verification for Outsourcing Database in Cloud"}

\author[1]{\fnm{Chebrolu Deepak} \sur{Kumar}}\email{deepakkumar.c19@iiits.in}

\author[1]{\fnm{Lilly Kumari} \sur{Biswas}}\email{	lillykumari.b19@iiits.in}

\author[1]{\fnm{Srijanee} \sur{Mookherji}}\email{srijanee.mookherji@iiits.in}

\author[1]{\fnm{Gowri Raghavendra Narayan} \sur{Kurmala}}\email{narayan.kgr@iiits.in}

\author [1]{\fnm{Vanga} \sur{Odelu}}\email{odelu.vanga@gmail.com}

\affil [1]{\orgdiv{Computer Science and Engineering}, \orgname{Indian Institute of Information Technology Sri City, Chittoor}, \city{Chittoor}, \postcode{517646}, \state{Andhra Pradesh}, \country{India}}




\abstract
{In 2022, Neela-Kavitha [Wireless Personal Communications, 123(3):2431–2448, 2022] proposed an improved RSA encryption algorithm (IREA) for cloud environment. In this paper, we review and comment on the correctness of the IREA technique. We prove that the private key generation in the proposed IREA is mathematical incorrect. That is, decryption of the cipher generated by the public key is not possible with the private key. Further, we discuss the possible modifications in IREA to make it correct decryption.} 


\maketitle

\section{Introduction}
Cloud Computing (CC) is a centralized system used to store sensitive data like financial details, healthcare data, defence, education, and so on. If such data is compromised, it can cause severe security breach. Thus, security is one of the most crucial challenge that needs to be handled in a cloud computing system. Service providers must guarantee safe and secure storage of information. Additionally, service providers must to protect themselves as well as their users from any form of security breaches. In order to maintain a secure system, cryptographic techniques are utilized in various researches~\cite{odelu2017,swathi2018}. Since the conventional RSA Algorithm~\cite{rivest1978method} is slow and directly can not be used for the cloud framework~\cite{neela2022improved}, Neela-Kavitha~\cite{neela2022improved} proposed an improved RSA encryption algorithm (IREA) for cloud framework. But, in this paper, we review IREA and present a comment on the correctness of algorithm.

\par
\textbf{Organization of the paper:} The paper is further organised as follows: In Section~\ref{s:prelim} we discuss the necessary mathematical preliminaries for understanding the review and analysis of IREA. In Section~\ref{s:review} we review Neela-Kavitha's protocol~\cite{neela2022improved}. Next, in Section~\ref{s:comment} we presented the correctness error in the protocol. In Section~\ref{s:modification} we provided a possible modification in IREA and finally we concluded in Section~\ref{s:conclusion}.

\section{Preliminaries} \label{s:prelim}
In this section, we present the required preliminaries to review and analyse Neela-Kavitha's~\cite{neela2022improved} Algorithm. 

\begin{definition}
    Fermat's Little Theorem~\cite{niven1991introduction}: Let $p$ be a prime number, and $a$ be an integer such that $a$ is not divisible by $p$ then, $a^{p-1} \equiv 1 \pmod {p}$.
\end{definition}

\begin{definition}
    Euler's Theorem~\cite{niven1991introduction}:  If $m$ is a positive integer and $a, n$ are coprimes, then $a \phi {(m)} \equiv 1 \pmod {m}$ where $\phi{(m)}$ is the Euler's totient function.
    Euler's totient function $\phi{(m)}$ of a positive integer $m > 1$ is defined to be the number of positive integers less than $m$ that are coprime to $m$.
\end{definition}

The convention RSA algorithm~\cite{rivest1978method} is as follows:
\begin{itemize}
\item \textbf{Setup:} Choose primes $b$ and $v$, and compute modulus $j = b \times v$ and Euler's totient $\phi{(j)} = (b-1) \times (v-1)$.
\item \textbf{Key Generation:} Choose $e$ such that $gcd(e,\phi{(j)}) = 1$. Then, $de \equiv 1 \pmod{\phi{(j)}}$ \\ The public key is $\{e,j\}$ and private key is $\{d,j\}$
\item \textbf{Encryption $E(\cdot)$: $E = M^{e} \pmod{j}$}, where $M$ is the plaintext message.
\item \textbf{Decryption $D(\cdot)$: $D = E^{d} \pmod{j}$}, where $E$ is the cipher text message. 
\end{itemize}

\section{Review of Neela-Kavitha's~\cite{neela2022improved} Algorithm} \label{s:review}
We review the Neela-Kavitha's~\cite{neela2022improved} proposed improved RSA algorithm IREA as follows:
\begin{itemize}
\item \textbf{IREA Key Generation:}
In this phase, the public key $p$ and the private key $d$ are generated. The steps for generating the key pairs are as follows:
\begin{itemize}
\item In the first step, two large prime numbers $b$ and $v$ are chosen and Modulus ${j}$ is computed, where $j = b \times v$.  
\item Next, $\phi{(j)}$ is computes as $\phi{(j)} = {b - 1}\times (v - 1)$. 
\item Then, a value $e$ is selected such that, $b < e <\phi{(j)}$ and is co-prime to $\phi{(j)}$ and $j$. 
\item The value $p$ is calculated as $p=(e\times 2) +1$ which serves as the public key. 
\item The private key $d$ is generated from the equation $d\times e \pmod{j} = 1$ 
\item Thus, public key is $(p, a)$ and private key is $(d, a)$ where, $a=j-1$. 
\end{itemize}
\item \textbf{IREA Encryption:} 
In this phase a message $M$ is encrypted using public key $p$ and modulus $j$ in the following way:
\begin{itemize}
\item The message $M$ is encrypted as $E$ using, $E=M^{(p-1)/2} \pmod{j}$
\end{itemize}
\item \textbf{IREA Decryption:}
In this phase a the encrypted cipher text $E$ is decrypted using private key $d$ and modulus $j$ in the following way:
\begin{itemize}
\item The cipher text $E$ is decrypted as $D = E^d \pmod{j}$ to get back the original plaintext message $M$. If the protocol functions correctly, $D = M$.
\end{itemize}

In this paper, the authors have presented a numerical example for IREA, the example is as follows:

\item \textbf{Numerical Example:} The authors chose two primes $b=5, v=11$ and the value $e = 13$. The complete example is demonstrated in Table~\ref{tab:numexp_base}.

\begin{table}[h]
\caption{IREA Numerical Example used in Neela-Kavitha~\cite{neela2022improved}}
\label{tab:numexp_base}
\begin{tabular}{|l|l|} \hline
\multicolumn{2}{|l|}{Key Generation Phase:} \\ \hline
\multicolumn{2}{|l|}{Consider two primes $b=5, v=11$ and calculate $ j = 5\times 11 = 55$,  $\phi{(j)} = 4\times 10 = 40$, a = j=1 = 54} \\
\multicolumn{2}{|l|}{Choosing $e = 13$, such that $b<e<\phi{(j)}$ and calculating public key $p = (e\times 2)+1 = (13\times 2)+1 = 27$} \\
\multicolumn{2}{|l|}{Finding $d$ such that $d\times e \pmod j =1$, $d\times 13 \pmod {55}$, $d = 17$} \\
\multicolumn{2}{|l|}{Public key = \{p,a\} = \{27,54\} and Private key = \{d,a\} = \{17,54\}} \\ \hline
Encryption Phase: & Decryption Phase: \\ \hline
Considering message $M=4$ & Considering Cipher Text $E=9$ \\
$E=M^{(p-1)/2} \pmod{a+1}$, $E=4^{13} \pmod{55}$ = 9 & $D = E^d \pmod{a +1}$, $D = 9^{17} \pmod{55}$ = 4 \\ \hline 
\multicolumn{2}{|c|}{The decrypted message $D$ is equal to the original message $M$} \\ \hline
\end{tabular}
\end{table}

\end{itemize}

\section{Comment on Neela-Kavitha's~\cite{neela2022improved} Scheme} \label{s:comment}

The proposed protocol IREA suffers from correctness error. According to the Euler and Fermat's little theorem~\cite{niven1991introduction}, $M$ is relatively prime to $j$, where $j=b\times v$ and $b$ and $v$ are large primes. 
\begin{equation} \label{e:euler_non_prime}
M^{(\phi(j))} \equiv 1 \pmod{j}.
\end{equation}
Here, $\phi{(j)}$ is the Euler's Totient function which gives the number of positive integers less than $j$ which are relatively prime to $j$. 
For prime numbers, $b$: 
\begin{equation} \label{e:euler_prime}
\phi (b) = b - 1
\end{equation}
By the elementary properties of the Euler's totient function:
\begin{equation} \label{e:phi_divides}
\phi(j) = \phi(b) \times  \phi(v) = ( b - 1 ) \times  ( v - 1) = j - ( b + v ) + 1.
\end{equation}
Since $d$ is relatively prime to $\phi(j)$, it has a multiplicative inverse $e$ in the ring of integers modulo $\phi(j)$ such that:
\begin{equation} \label{e:ed}
e \times  d \equiv 1 \pmod{\phi(j)}.
\end{equation}
Therefore, for some integer $k$,
\begin{equation} \label{e:ed_phi}
M^{ed} \equiv M^{k\times \phi(j)+1} \pmod {j}
\end{equation}
From Equations~\ref{e:euler_non_prime} and~\ref{e:euler_prime} we get, 
\begin{equation}
M^{b-1} \equiv 1 \pmod b 
\end{equation}
and since $(b-1)$ divides $\phi{(j)}$ as per Equation~\ref{e:phi_divides}, we can say, 
\begin{equation} \label{e:phi_m_b}
M^{k\times \phi(j) + 1} \equiv M \pmod b
\end{equation}
This equality hold for all $M$, thus, 
\begin{equation} \label{e:phi_m_v}
M^{k\times \phi(j) + 1} \equiv M \pmod v
\end{equation}
Therefore, from Equation~\ref{e:ed_phi} and~\ref{e:phi_m_b} we can obtain,
\begin{equation}
M^{ed} \equiv M^{k\times \phi(j)+1} \pmod {j} \equiv M \pmod {j}
\end{equation}

Thus, in the proposed protocol, as per Equation~\ref{e:ed}, $d\times e \pmod j \equiv 1$, where $j=b\times v$, provides wrong result causing mismatched decryption of the message. 

To support our claim we demonstarate the same numerical example with a changed value of $e$. The example is presented in Table~\ref{tab:numexp_wrong}.

\begin{table}[h]
\caption{IREA Numerical Example using Neela-Kavitha et al.'s protocol with $e=7$ }
\label{tab:numexp_wrong}
\begin{tabular}{|l|l|} \hline
\multicolumn{2}{|l|}{Key Generation Phase:} \\ \hline
\multicolumn{2}{|l|}{Consider two primes $b=5, v=11$ and calculate $ j = 5\times 11 = 55$,  $\phi{(j)} = 4\times 10 = 40$, a = j=1 = 54} \\
\multicolumn{2}{|l|}{Choosing $e = 7$, such that $b<e<\phi{(j)}$ and calculating public key $p = (e\times 2)+1 = (7\times 2)+1 = 15$} \\
\multicolumn{2}{|l|}{Finding $d$ such that $d\times e \pmod j =1$, $d\times 7 \pmod {55}$, $d = 8$} \\
\multicolumn{2}{|l|}{Public key = $\{p,a\} = \{15,54\}$ and Private key = $\{d,a\} = \{8,54\}$} \\ \hline
Encryption Phase: & Decryption Phase: \\ \hline
Considering message $M=4$ & Considering Cipher Text $E=49$ \\
$E=M^{(p-1)/2} \pmod{a+1}$, $E=4^{7} \pmod{55}$ = 49 & $D = E^d \pmod{a +1}$, $D = 49^{8} \pmod{55}$ = 26 \\ \hline 
\multicolumn{2}{|c|}{The decrypted message $D = 26$ is not equal to the original message $M=4$} \\ \hline
\end{tabular}
\end{table}

\section{Possible Modification} \label{s:modification}

In IREA, using the Euler's and Fermat's Little Theorem, the private key generation step should be changed as $d\times e \pmod{\phi{(j)}} = 1$. Thus the protocol changes as follows:

\begin{itemize}
\item \textbf{IREA Key Generation:}
In this phase, the public key $p$ and the private key $d$ are generated. The steps for generating the key pairs are as follows:
\begin{itemize}
\item In the first step, two large prime numbers $b$ and $v$ are chosen and Modulus ${j}$ is computed, where $j = b \times v$.  
\item Next, $\phi{(j)}$ is computes as $\phi{(j)} = {b - 1}\times (v - 1)$. 
\item Then, a value $e$ is selected such that, $b < e <\phi{(j)}$ and is co-prime to $\phi{(j)}$ and $j$. 
\item The value $p$ is calculated as $p=(e\times 2) +1$ which serves as the public key. 
\item The private key $d$ is generated from the equation $d\times e \pmod{\phi{(j)}} = 1$ 
\item Thus, Public key is $(p, a)$ and Private key is $(d, a)$ where, $a=j-1$. 
\end{itemize}
\item \textbf{IREA Encryption:} 
In this phase a message $M$ is encrypted using public key $p$ and modulus $j$ in the following way:
\begin{itemize}
\item The message $M$ is encrypted as $E$ using, $E=M^{(p-1)/2} \pmod{j}$
\end{itemize}
\item \textbf{IREA Decryption:}
In this phase a the encrypted cipher text $E$ is decrypted using private key $d$ and modulus $j$ in the following way:
\begin{itemize}
\item The cipher text $E$ is decrypted as $D = E^d \pmod{j}$ to get back the original plaintext message $M$. If the protocol functions correctly, $D = M$.
\end{itemize}
\end{itemize}

To prove the correctness of the above discussed protocol, we present a numerical example using the same value that failed for Neela-Kavitha's~\cite{neela2022improved} protocol. The example is presented in the Table~\ref{tab:numexp_correct}.

\begin{table}[h]
\caption{Modified IREA Numerical Example with $e=7$ }
\label{tab:numexp_correct}
\begin{tabular}{|l|l|} \hline
\multicolumn{2}{|l|}{Key Generation Phase:} \\ \hline
\multicolumn{2}{|l|}{Consider two primes $b=5, v=11$ and calculate $ j = 5\times 11 = 55$,  $\phi{(j)} = 4\times 10 = 40$, a = j=1 = 54} \\
\multicolumn{2}{|l|}{Choosing $e = 7$, such that $b<e<\phi{(j)}$ and calculating public key $p = (e\times 2)+1 = (7\times 2)+1 = 15$} \\
\multicolumn{2}{|l|}{Finding $d$ such that $d\times e \pmod j =1$, $d\times 7 \pmod {40}$, $d = 23$} \\
\multicolumn{2}{|l|}{Public key = $\{p,a\} = \{15,54\}$ and Private key = $\{d,a\} = \{23,54\}$} \\ \hline
Encryption Phase: & Decryption Phase: \\ \hline
Considering message $M=4$ & Considering Cipher Text $E=49$ \\
$E=M^{(p-1)/2} \pmod{a+1}$, $E=4^{7} \pmod{55}$ = 49 & $D = E^d \pmod{a +1}$, $D = 49^{23} \pmod{55}$ = 4 \\ \hline 
\multicolumn{2}{|c|}{The decrypted message $D = 4$ is equal to the original message $M=4$} \\ \hline
\end{tabular}
\end{table}

The decryption phase functions as expected and the original message, ``M = 4" is obtained. Thus, correctness of the modified protocol is proven. 

\section{Conclusion} \label{s:conclusion}
In this paper, we review Neela-Kavitha's improved RSA technique. The proposed protocol suffers from mathematically inaccuracy. We present numerical example to support the claim. Additionally, we presented a suggested modification and proved the correctness of the modified protocol using numerical values.

\bibliography{WirelessComm}

\end{document}